\documentclass[12pt,showpacs,showkeys,superscriptaddress,aps]{revtex4}
\usepackage{amsmath,amsthm,amssymb}
\usepackage{mathrsfs}

\usepackage{fancyhdr}
\usepackage{ascmac}

\usepackage{epic, eepic}
\usepackage{graphicx} 
\usepackage{here}

\usepackage{braket}
\usepackage{bm}

\usepackage{url}

\usepackage{color}

\newcommand{\maprightu}[1]{%
\smash{\mathop{%
\hbox to 1cm{\rightarrowfill}}\limits^{#1}}}
\newcommand{\maprightd}[1]{%
\smash{\mathop{%
\hbox to 1cm{\rightarrowfill}}\limits_{#1}}}
\newcommand{\mapleftu}[1]{%
\smash{\mathop{%
\hbox to 1cm{\leftarrowfill}}\limits^{#1}}}
\newcommand{\mapleftd}[1]{%
\smash{\mathop{%
\hbox to 1cm{\leftarrowfill}}\limits_{#1}}}
\newcommand{\mapnerssss}[1]{%
\smash{\mathop{%
\hbox to 3cm{\nearrow}}\limits^{#1}}}

\begin{document}
\title{Derivation of Higher Dimensional Periodic Recurrence Equations \\
by Nested Structure of Complex Numbers\\
}
\author{Tsukasa Yumibayashi}
\email[email : ]{tsukasa.yumibayashi@otsuma.ac.jp}
\affiliation{Department of Social Information Studies, Otsuma Women's University,\\
Karakida 2-7-1, Tama city, Tokyo, 192-0397 Japan}

%%%%%%%%%%%%%%%%%%%%%%%%%%%%%%
%%%%%%%%%%%%%%%%%%%%%%%%%%%%%%
%%%%%%%%%%%%%%%%%%%%%%%%%%%%%%
\begin{abstract}
In this paper, we give a procedure for derivation of higher dimensional periodic recurrence equations(PREs) by nested structure of complex numbers. 
\end{abstract}
%\pacs{}
%%%%%%%%%%%%%%%%%%%%%%%%%%%%%%
\maketitle
%%%%%%%%%%%%%%%%%%%%%%%%%%%%%%
\section{Introduction}
%Philosophy
What is the time? When people want to measure the time, people use a clock. All clocks utilize a ``periodic'' motion, for example, many clocks often have a structure of harmonic oscillator which has the same period, {\it i.e.} frequency, for all initial points. Hence almost all people interpret the time as the number of times of an initial point coming back to the initial point. In particular, the harmonic oscillator is an object in continuum dynamical systems. In this paper, we want to investigate about periodic motion in discrete dynamical systems.

%PRE
For a map $F$ and a fixed number $n \in \mathbb{N}$, the $F$ is called a periodic recurrence equation(PRE) of period $n$, when for all initial points of the $F$ are $n$ periodic points\cite{RE0}\cite{RE1}. There is similar argument \cite{GPM1} from different point of view.

For examples\cite{RE0}:
\begin{itemize}
\item PRE of period 2(dim 1):
\begin{equation}
X=\frac{a}{x}, \quad a \neq 0, \quad x, X, a \in \mathbb{C},
\label{2pre}
\end{equation}
\item PRE of period 5(dim 2):
\begin{equation}
X=\frac{1+x}{y}, \quad Y=x, \quad x, y, X, Y \in \mathbb{C}.
\label{5pre}
\end{equation}
\end{itemize}
We can check the periodicity for these maps: all initial points, all orbits can be explicitly written in ``symbolic'' calculus as follows:
\begin{itemize}
\item PRE of period 2:
\begin{equation}
x \rightarrow \frac{a}{x} \rightarrow \frac{a}{\frac{a}{x}}=x.
\end{equation}
\item PRE of period 5:
\begin{eqnarray}
(x, y) &\rightarrow& \left(\frac{1+x}{y}, x \right) \rightarrow \left( \frac{1+x+y}{xy}, \frac{1+x}{y} \right) \rightarrow \nonumber \\
&\rightarrow& \left( \frac{1+y}{x}, \frac{1+x+y}{xy} \right) \rightarrow \left( y, \frac{1+y}{x} \right) \rightarrow \left( x, y \right).
\end{eqnarray}
\end{itemize}
Here, ``symbolic'' means ``don't touch any information of initial points''. This fact is the key point of the present paper.

A PRE can be interpreted as a ``clock'' in discrete dynamical systems. In particular, a set of PRE is a ``basis'' for an integrable system, like a set of harmonic oscillators is a basis for periodic function {\it i.e.} Fourier basis. The meaning of this will be discussed in conclusion.

%Motivation
Until recently, well known examples of PRE are mostly low dimensional cases\cite{RE0}\cite{RE1}\cite{GPM1}. However, in \cite{RE2}, authors give a procedure for derivation of infinitely many higher dimensional cases using the structure of invariant variety of periodic points(IVPP)\cite{IVPPa}\cite{IVPPb}\cite{IVPPbook}. Here, IVPP is a variety of periodic points which cover some level sets. In this paper, we would give a new procedure for derivation of infinitely many higher dimensional PREs from a low dimensional PRE and/or a map which has IVPP using by symbolic structure of PRE.

%Abstract1
In this paper, our discussion is based on $d$ dimensional complex rational maps
\begin{equation}
F:\bm{x}:=(x_1,\dots, x_d) \mapsto \bm{X}:=(X_1, \dots, X_d), \quad \bm{x} \in \mathbb{C}^d, \quad \bm{X} \in \mathbb{C}(\bm{x})^d, \label{rm}
\end{equation}
where $\mathbb{C}(\bm{x})$ is the rational function field over $\mathbb{C}$.

For studying a concrete map like (\ref{rm}), we often use some computer softwares, for examples, Mathematica, Maple etc..., and when we are carrying out calculations and writing graphs for finding some structures of the map as complex map on these softwares, we must rewrite an explicit form which replaces each variable $x_i$ by $x_i+ \sqrt{-1}y_i$ for $i=1,\dots, d$. This replacement is the first step for our procedure.

Furthermore, the complex number often interpreted as a 2 dimensional vector over $\mathbb{R}$, and then the map (\ref{rm}) also can be interpreted as a $2d$ dimensional REAL map $F[1]:\mathbb{R}^{2d}\rightarrow \mathbb{R}^{2d}$ which the symbol ``$[1]$'' means one step after our procedure for the map (\ref{rm}). Here, once again, we replace the real variables in $F[1]$ by complex variables, therefore we get a new $2d$ dimensional COMPLEX map $F[1]:\mathbb{C}^{2d}\rightarrow \mathbb{C}^{2d}$. These replacements are the second step for our procedure.

These steps can be iterated $n$ times for any $n \in \mathbb{N}$. Hence we can get a $2^nd$ dimensional complex map $F[n]$ for any $n \in \mathbb{N}$. The sequence of higher dimensional maps $\{F[n] \}_{n\in \mathbb{N}}$ is our main implement.

A PRE provides periodic motion for symbolic calculus, then the PRE conserves of this periodic structure by taking our procedure. 
It means that variables of a PRE are not related in neither abstract notation nor explicit notation. For example, when variables are written $x$ or $x+\sqrt{(-1)}y$, the PRE provides periodic motion of the same period.

%Outline
Before finishing this section, we give the outline of this paper: In \S 2, we introduce about IVPP for derivation of RPEs, in \S 3, we show our procedure, in \S 4, we give some examples, and in \S 5 we give conclusion.

%%%%%%%%%%%%%%%%%%%%%%%%%%%%%%
\section{IVPP and Derivation of IVPPs}
In this section, we give a review about invariant variety of periodic points(IVPP)\cite{IVPPa}\cite{IVPPb}\cite{IVPPbook} and derivation of IVPPs from singularity confinement(SC)\cite{ysw} to obtain PREs\cite{RE2}. 

\subsection{IVPP}
An IVPP of period $n$ of a $d$ dimensional rational map 
\begin{equation}
F : \bm{x} \mapsto \bm{X}, \quad \bm{x}, \bm{X} \in \mathbb{C}^d,
\label{rm1}
\end{equation}
with $p$ invariants 
\begin{equation}
\bm{r}: \mathbb{C}^d \rightarrow \mathbb{C}^p, \ \mathrm{s.t.} \ \bm{r}(\bm{x})=\bm{r}(\bm{X}), \quad r(\bm{x})\in \mathbb{C}(\bm{x}),
\label{inv1}
\end{equation}
is defined as a variety of periodic points of period $n$
\begin{equation}
\Set{ \bm{x} \in \mathbb{C}^d | (F^{(n)}(\bm{x})-\bm{x}=0) \ \wedge \ (F^{(m)}(\bm{x})-\bm{x} \neq 0, \ m \leq n) },
\end{equation}
which is given by only the invariants
\begin{equation}
\Set{ \bm{x} \in \mathbb{C}^d | \bm{\gamma}^{(n)}(\bm{r}(\bm{x}))=0 }, \quad \bm{\gamma^{(n)}} \circ \bm{r} : \mathbb{C}^d \rightarrow \mathbb{C}^{d-p}.
\label{IVPP}
\end{equation}

The IVPP has an important property which is called the IVPP theorem,

\bigskip

{\it \noindent Let F be a $d$ dimensional rational map with $p$ invariants. If $p \geq d/2$, an IVPP and discrete periodic points on a level set of any period do not exist in one map, simultaneously.}

A map (\ref{rm1}) can be restricted on a level set. Because, any initial points on a level set are confined on this level set. Moreover an IVPP of period $n$ is a level set and a set of $n$ periodic points. Therefore a map on an IVPP of period $n$ becomes a PRE of period $n$.

\subsection{Derivation of IVPPs and PREs from SC}
The IVPP also has an important phenomenon, that is the derivation of a sequence of IVPPs from SC\cite{SC1}\cite{SC2}, and then it is equivalent to the phenomenon which gives a sequence of PREs. First we give a simple example for the explanation of this phenomenon:

\begin{itemize}
\item 2 dimensional M\"obius map\cite{sshyw}:
\begin{equation}
X=x\frac{1-y}{1-x}, \quad Y=y\frac{1-x}{1-y}, \quad x, y, X, Y \in \mathbb{C}, \label{2mob}
\end{equation}
with an invariant,
\begin{equation}
r(x, y)=r(X, Y)= xy. \label{2mob-inv}
\end{equation}
\item Parametrize an initial point $p_0$ for SC by the invariant:
\begin{equation}
p_0 \in \{ (1-x=0) \wedge (R=xy) \} \quad \Rightarrow \quad p_0=(1, R), \quad R \in \mathbb{C}.
\label{ip}
\end{equation}
\item 
Make an orbit of iterations of the map (\ref{2mob}) for the initial point (\ref{ip}):
\begin{eqnarray}
p_0=(1, R) &\rightarrow& (\infty, 0) \rightarrow (-1, -R) \rightarrow \nonumber \\
&\rightarrow& \left(-\frac{1+R}{2}, -\frac{2R}{1+R} \right) \rightarrow \left( -\frac{1+3R}{3+R}, -\frac{R(3+R)}{1+3R} \right) \rightarrow \dots.
\end{eqnarray}
Notice that $p_0$ goes back to a finite point keeping information of $p_0$, this phenomenon is called SC\cite{SC1}\cite{SC2}.
\item Take ``a condition for the point going back to infinity after $n$ iterations of the map, again''. This condition is equivalent to the condition of IVPP of period $n$:
\begin{eqnarray}
\gamma^{(2)}(R) &=& none, \nonumber \\
\gamma^{(3)}(R) &=& 3+R, \quad \mathrm{so \ \ on}.
\end{eqnarray}

\item Restrict the map (\ref{2mob}) on the IVPP of period $n$. This map is a PRE of period $n$:
\begin{equation}
X=\frac{x+3}{1-x}, \quad \mathrm{so \ \ on}.
\label{2mob3}
\end{equation}
CHECK:
\begin{equation}
x \rightarrow \frac{x+3}{1-x} \rightarrow \frac{x-3}{1-x} \rightarrow x.
\end{equation}
\end{itemize}

More generally, the algorithm of derivation of IVPPs and PREs from SC is as follows:
\begin{enumerate}
\item Solve the $d$ conditions for parametrizing the initial point $\bm{x}^{0}$ by the invariants:
\begin{eqnarray}
D_1^{(j)}(\bm{x})&=&0,\quad j=1,2,...,d-p, \nonumber \\
\bm{r}(\bm{x})&=& \bm{R}, \quad \bm{R} \in \mathbb{C}^p,
\label{conditions}
\end{eqnarray}
where $D_1^{(j)}(\bm{x}), j=1,2,...,d-p$ are the denominators of $j$ iterated map of the component $1$. 
\item Compute $F^{(n)}(\bm{x}^{0}),\  n\ge N_{sc}$ iteratively to find $D_1^{(n)}(\bm{x}^{0})$, which is a polynomial function of the invariants, and where ``SC step'' $N_{sc}$ is the number of iterations necessary to go back to a finite point. 
\item Take $d-p$ irreducible polynomials, one from each of the set
\[
\left\{D_1^{(n+1)}(\bm{x}^{0}),\ D_1^{(n+2)}(\bm{x}^{0}),\ ...,\ D_1^{(n+d-p)}(\bm{x}^{0})\right\},\qquad n\ge N_{sc}-1.
\] 
If the polynomials in this set are all independent, the intersection of the elements is a set of periodic points of period $n$.

\item Restrict the map on the IVPP of period $n$, then we get a $d-p$ dimensional PRE of period $n$.
\end{enumerate}

%%%%%%%%%%%%%%%%%%%%%%%%%%%%%%
\section{Procedure}
In this section, we give our procedure for the derivation of a sequence of higher dimensional maps by some base maps.

Let $F$ be a $d$ dimensional ``complex'' map (\ref{rm1})
$$
F:(x_1, \dots, x_d) \mapsto (X_1, \dots, X_d), \quad x_i, X_i \in \mathbb{C}, \quad i=1,\dots, d,
$$
with $p$ ``complex'' invariants (\ref{inv1})
$$
r_j:= r_j (x_1, \dots, x_d)= r_j (X_1, \dots, X_d), \quad r_j \in \mathbb{C}, \quad j=1,\dots, p.
$$

The map $F$ is $d$ dimensional complex map and $r_j, j=1, \dots, p$ are $p$ complex invariants, however it is not clear in our representation (\ref{rm1}), (\ref{inv1}). Now we define the complexified map
\begin{equation}
A^{(1)}_i : x_i \mapsto x_i + \sqrt{-1} x_{d+i}, \quad B^{(1)}_j : r_j \mapsto r_j + \sqrt{-1} r_{p+j}, \quad i =1, \dots, d, \quad j=1, \dots, p, \label{cm}
\end{equation}
to give an explicit form. Therefore we can rewrite the map $F$ to an explicit form as follows
\begin{equation}
F[1]: (x_1, \dots, x_d, x_{d+1}, \dots, x_{2d}) \mapsto (X_1, \dots, X_d, X_{d+1}, \dots, X_{2d}),
\end{equation}
where
\begin{equation}
X_i + \sqrt{-1} X_{d+i} = [F ( A^{(1)}_1 (x_1), \dots, A^{(1)}_d(x_d))]_i, \quad i=1,\dots, d, \label{com-map}
\end{equation}
{\it i.e. }
\begin{eqnarray}
X_i &=&  [F[1](x_1, \dots, x_{2d})]_i= \Re([F ( A^{(1)}_1 (x_1), \dots, A^{(1)}_d(x_d))]_i), \\
X_{d+i} &=&  [F[1](x_1, \dots, x_{2d})]_{d+i}= \Im( [F ( A^{(1)}_1 (x_1), \dots, A^{(1)}_d(x_d))]_i ), \quad \quad i=1,\dots, d.
\end{eqnarray}

Usually, we think that (\ref{com-map}) is $2d$ dimensional REAL map with $2p$ REAL invariants, {\it i.e.} $x_i, X_i \in \mathbb{R}^d, i=1,\dots, 2d$, and $r_j \in \mathbb{R}, j=1,\dots, 2p$.
Here we replace the real variables $(x_1, \dots, x_d, x_{d+1}, \dots, x_{2d})$ by complex variables. Therefore, of course we get a $2d$ dimensional complex map with $2d$ complex invariants. 

Furthermore we can ITERATE this procedure for any times! Therefore we can get a $2nd$ dimensional complex map with $2np$ complex invariants for any $n \in \mathbb{N}$
\begin{eqnarray}
X_i &=& \Re([F[n] ( A^{(n)}_1 (x_1), \dots, A^{(n)}_{nd} (x_{nd}))]_i), \\
X_{2nd+i} &=& \Im([F[n] ( A^{(n)}_1 (x_1), \dots, A^{(n)}_{nd} (x_{nd}))]_i), \quad i=1,\dots, 2nd,
\end{eqnarray}
by $n$-th complexified map,
\begin{equation}
A^{(n)}_i : x_i \mapsto x_i + \sqrt{-1} x_{nd+i}, \quad B^{(n)}_j : r_j \mapsto r_j + \sqrt{-1} r_{np+j}, \quad i=1, \dots, nd, j=1, \dots, np,
\end{equation}
where $F[n]$ is given by taking our procedure for $n$ steps.

Now we give small remark that if a map satisfies the condition $p \geq d/2$ for IVPP theorem, then this condition is preserved at $^\forall n \in \mathbb{N}$, because,
\begin{equation}
p \geq d /2 \Rightarrow 2np \geq 2nd/2.
\end{equation}

However, a difference of dimension $2nd$ and the number of invariants $2np$ is increased by $d-p$ at each step, hence, we must use the generalized algorithm of derivations of IVPPs from SC when we calculate the IVPPs for the map $F[n]$.

%%%%%%%%%%%%%%%%%%%%%%%%%%%%%%
\section{Example}
In this section, we give some examples of our procedure.

%%%%%%%%%%%%%%%%%%%%%%%%%%%%%%
\subsection{PRE of Period 2}
The base map is the PRE of period 2 (\ref{2pre})
$$
X=\frac{1}{x}.
$$
Then we get a ``doublet map'' as follows
\begin{equation}
X=\frac{x}{x^2+y^2}, Y=-\frac{y}{x^2+y^2}, \label{22}
\end{equation}
by the doublet variables $x \mapsto x+\sqrt{-1}y$.

And then we get a ``quadruplet map'' as follows
\begin{eqnarray}
X &=& \frac{x(x^2+y^2+u^2-v^2)+2yuv}{((x+v)^2+(y-u)^2)((x-v)^2+(y+u)^2)}, \nonumber \\
U &=& -\frac{u(x^2-y^2+u^2+v^2)+2xyv}{((x+v)^2+(y-u)^2)((x-v)^2+(y+u)^2)}, \nonumber \\
Y &=& -\frac{y(x^2+y^2-u^2+v^2)+2xuv}{((x+v)^2+(y-u)^2)((x-v)^2+(y+u)^2)}, \nonumber \\
V &=& \frac{v(-x^2+y^2+u^2+v^2)+2yuv}{((x+v)^2+(y-u)^2)((x-v)^2+(y+u)^2)}, \label{24}
\end{eqnarray}
by the quadruplet variables $x \mapsto x+\sqrt{-1}u, y \mapsto y+\sqrt{-1}v$.

Of course, (\ref{22}) and (\ref{24}) are PRE of period 2.

%%%%%%%%%%%%%%%%%%%%%%%%%%%%%%
\subsection{PRE of Period 5}
The base map is the PRE of period 5 (\ref{5pre})
$$
X=\frac{1+x}{y}, \quad Y=x. 
$$
Then we get a ``doublet map'' as follows
\begin{eqnarray}
X &=& \frac{y(1+x)+uv}{y^2+v^2}, \nonumber \\
U &=& \frac{-v(1+x)+yu}{y^2+v^2}, \nonumber \\
Y &=& x, \nonumber \\
V &=& u, \label{54}
\end{eqnarray}
by the doublet variables $x \mapsto x+\sqrt{-1}u, y \mapsto y+\sqrt{-1}v$.

And the orbit of the map (\ref{54}) is
\begin{eqnarray}
(x, u, y, v) &\rightarrow& \left(\frac{y(1+x)+uv}{y^2+v^2}, \frac{-v(1+x)+yu}{y^2+v^2}, x, u \right) \rightarrow \nonumber \\
&\rightarrow& \left( \frac{xy(1+x+y)+xv^2+yu^2-uv}{(x^2+u^2)(y^2+v^2)}, -\frac{xv(1+x)+yu(1+y)+uv(u+v)}{(x^2+u^2)(y^2+v^2)}, \right. \nonumber \\
&& \quad \left. \frac{y(1+x)+uv}{y^2+v^2}, \frac{-v(1+x)+yu}{y^2+v^2} \right) \rightarrow \nonumber \\
&\rightarrow& \left(\frac{x(1+y)+uv}{x^2+u^2}, -\frac{u(1+y)-xv}{x^2+u^2}, \right. \nonumber \\
&& \left. \frac{xy(1+x+y)+xv^2+yu^2-uv}{(x^2+u^2)(y^2+v^2)}, \frac{xv(1+x)+yu(1+y)+uv(u+y)}{(x^2+u^2)(y^2+v^2)} \right) \rightarrow \nonumber \\
&\rightarrow& \left( y, v, \frac{x(1+y)+uv}{x^2+u^2}, -\frac{u(1+y)-xv}{x^2+u^2} \right) \rightarrow \left( x, u, y, v \right).
\end{eqnarray}
Therefore we could check the map (\ref{54}) is a PRE of period 5.

%%%%%%%%%%%%%%%%%%%%%%%%%%%%%%
\subsection{2 dimensional M\"obius Map}
The base map is the 2 dimensional M\"obius map (\ref{2mob})
$$
X=x\frac{1-y}{1-x}, \quad Y=y\frac{1-x}{1-y},
$$
with the invariant (\ref{2mob-inv})
$$
r(x,y)=xy.
$$

\bigskip

We know general formula of the IVPPs (\ref{IVPP}) of the map (\ref{2mob}) as follows\cite{sshyw}:
\begin{equation}
\gamma^{(n)}(r) =r+ \tan^2 \left( \frac{\pi m}{n} \right), \quad m= 1,2 \dots, n-1, \quad n=3,4, \dots,
\end{equation}
and explicit forms, which will be discussed in this paper, as follows:
\begin{eqnarray}
\gamma^{(2)}(r) &=& none, \nonumber \\
\gamma^{(3)}(r) &=& 3+r, \nonumber \\
\gamma^{(4)}(r) &=& 1+r, \quad \mathrm{so \ \ on.} \nonumber
\end{eqnarray}
\begin{center}
\begin{figure}[H]
\begin{center}
\includegraphics[scale=0.35, bb=0 0 549 563]{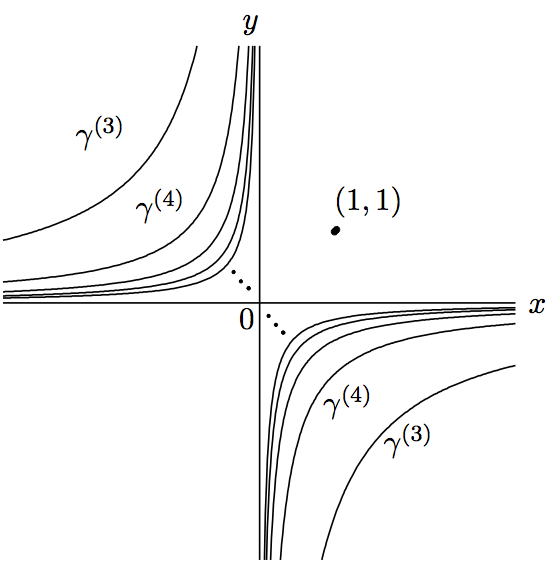}
\caption{IVPPs of the map (\ref{2mob})}
\end{center}
\end{figure}
\end{center}

\bigskip

Then we get ``doublet map'' as follows
\begin{eqnarray}
X &=& \frac{x(1-x)(1-y)-u^2(1-y)+uv}{(1-x)^2+u^2}, \nonumber \\
U &=& \frac{-xv(1-x)+u(1-y+uv)}{(1-x)^2+u^2}, \nonumber \\
Y &=& \frac{y(1-y)(1-x)-v^2(1-x)+uv}{(1-y)^2+v^2}, \nonumber \\
V &=& \frac{-yu(1-y)+v(1-x+uv)}{(1-y)^2+v^2},
\label{dmap}
\end{eqnarray}
with doublet invariants
\begin{equation}
r(x,u,y,v)=xy-uv, \quad s(x,u,y,v)=xv+yu,
\end{equation}
by the doublet variables $x \mapsto x+\sqrt{-1}u, y \mapsto y+\sqrt{-1}v$.

\bigskip

Here we need the IVPP for the map (\ref{dmap}) 
for to get REs. We use a method of derivations of IVPPs from SC:
\begin{enumerate}
\item Parametrize an initial point for SC:
\begin{eqnarray}
p_0 &=& \{ (D_1^{(1)}(x,u,y,v)=0) \wedge (D_1^{(2)}(x,u,y,v)=0) \nonumber \\
&& \quad \quad \wedge (r(x,u,y,v)=R) \wedge (s(x,u,y,v)=S) \} \nonumber \\
&\Rightarrow& p_0 = \left(\frac{1}{4}(R\pm \sqrt{-1}S), -\frac{1}{4}(-S\pm(R-1)), \right. \nonumber \\
&& \quad \left. \frac{\pm(R^2-3R-S^2)+\sqrt{-1}S}{\pm(1+R)+\sqrt{-1}S},  \frac{\pm3S+\sqrt{-1}(R^2-R+S^2)}{2(\pm(1+R)+\sqrt{-1}S)} \right).
\end{eqnarray}
\item Make the flow of iterations of the map(for short, we write only component of $x$):
\begin{eqnarray}
p_0 &\rightarrow& \infty \rightarrow \infty \rightarrow \frac{1}{4}(-R-3\pm\sqrt{-1}S) \rightarrow -\frac{\mp(R^2+10R+5+S^2)+4\sqrt{-1}S}{4(\mp(R+3)+\sqrt{-1}S)} \nonumber \\
&\rightarrow& \frac{R(R^2+21R+35)+S^2(15+R)\mp \sqrt{-1}S(R^2+6R+9+S^2)}{8(\pm(R+3)+\sqrt{-1}S)(\mp(R+1)+\sqrt{-1}S)} \rightarrow \dots.
\end{eqnarray}
\item The conditions
\begin{equation}
(\mp(R+3)+\sqrt{-1}S=0) \wedge (\mp(R+1)+\sqrt{-1}S=0), \quad \mathrm{so \ \ on,}
\label{ivpp12}
\end{equation}
give IVPP and so on.
\end{enumerate}

The IVPP (\ref{ivpp12}) has period 12, because if we drop the imaginary part {\it i.e.} when $S=0$, then the condition (\ref{ivpp12}) becomes common part of IVPP of period 3 and 4. And then, we get the PRE of period 12 which is restricted the map (\ref{dmap}) on the IVPP (\ref{ivpp12}) as follows
\begin{eqnarray}
X&=&\frac{L+x-2+\sqrt{-1}u}{M_+ M_-}, \nonumber \\
U&=&\frac{-3u+\sqrt{-1}(x-1)}{M_+ M_-},
\label{mob12}
\end{eqnarray}
where $M_{\pm}=-\sqrt{-1}(1-x)\pm u$, $L=x^2+u^2$ and $(R, S)=(-2, -\sqrt{-1})$ for easy to write. We give a flow of the (\ref{mob12}) in appendix.

When the doubled map (\ref{dmap}) is considered at $u=v=0$, (\ref{dmap}) returns  to the base map (\ref{2mob}). However, when the map (\ref{mob12}) is considered at $u=0$, (\ref{mob12}) dose not return to a 1 dimensional PRE, hence, (\ref{mob12}) is independent PRE from the PRE just given by the base map.

%%%%%%%%%%%%%%%%%%%%%%%%%%%%%%
\section{Conclusion}
In conclusion, our procedure could find NEW infinitely many PREs from some base maps which are PREs and/or maps have IVPPs. In particular, last example, the 2 dimensional M\"obius map suggest some fact as follows: For a $d$ dimensional map $F$ with $p$ invariants, if $F$ has SC step $N_{sc}+(d-p)-1$ by the conditions that $(d-p)$ denominators are zero for derivation of IVPPs, then $F[n]$ has SC step $N_{sc}+2^n(d-p)-1$ by conditions that $2^n(d-p)$ denominators are zero for derivation of IVPPs. Therefore, we can get periodic points which have only greater than $2^n(d-p)$ period, and concrete period is given a formula as follows,
\begin{equation}
\prod_{k=m}^{m+2^n(d-p)-1} k, \quad ^\exists m \geq N_{sc}+2^n(d-p)-2.
\end{equation}

Furthermore, \cite{syw} suggested that a derivation of sequence of IVPPs from SC is interpreted as projective resolution of singularity of a map in a triangulated category. 
We reinterpret the fact by our results that a map can be expanded by a sequence of PREs like a Fourier expansion, {\it i.e.}, a map $\bm{X}= F(\bm{x})$ which has IVPP can be written as
$$
F |_{\mathrm{periodic \ points}} (\bm{x})= \sum_{n = 2}^{\infty} \delta_{0, \bm{\gamma}^{(n)}(\bm{r}(\bm{x}))} F_n(\bm{x}),
$$
where $F_n(\bm{x})$ is PRE of period $n$ which is to restrict on an IVPP of period $n$.

In this paper, the complexified map(\ref{cm}) is performed of our procedure. However, it does NOT need to restrict on complexified. Because our procedure uses only symbolical ``algebraic structure'' of the PRE. Hence, we need only algebraic structure which are included every operation for the PRE. Therefore, for example, if we don't need the commutativity, then we can use ``quaternion''. A most simplest example, which is given by transformation $x \rightarrow x+iq_1+jq_2+kq_3$ for the PRE of period 2 (\ref{2pre}) as follows,
$$
X=\frac{1}{x} \quad \Rightarrow \quad X=\frac{x}{r^2}, \quad Q_i=\frac{-q_i}{r^2}, \quad i=1\dots, 3,
$$
where $r^2=x^2+q_1^2+q_2^2+q_3^2$.

%%%%%%%%%%%%%%%%%%%%%%%%%%%%%%
\section*{Acknowledgement}
I would like to thank Dr. S. Saito and Mr. Y. Wakimoto for useful discussion.

%%%%%%%%%%%%%%%%%%%%%%%%%%%%%%

\section*{Appendix: The flow of (\ref{mob12})}

\begin{eqnarray}
(x, u) &\rightarrow& \left(\frac{L+x-2+\sqrt{-1}u}{M_+ M_-}, \frac{-3u+\sqrt{-1}(x-1)}{M_+ M_-} \right) \nonumber \\
&\rightarrow& \left(\frac{-L+4x+1+2\sqrt{-1}u}{2N_+ K}, \frac{-4u+\sqrt{-1}(-L+2x-1)}{2N_+ K} \right) \nonumber \\
&\rightarrow& \left(\frac{\sqrt{-1}(L+2x-1)}{2N_-}, \frac{-L-1+2\sqrt{-1}u}{2N_-} \right) \nonumber \\
&\rightarrow& \left(\frac{\sqrt{-1}(L-2x-3)}{2M_-}, \frac{L+3-2\sqrt{-1}u}{2M_-} \right) \nonumber \\
&\rightarrow& \left(\frac{3x-1+\sqrt{-1}u}{M_- N_+}, \frac{-u+\sqrt{-1}(-L+x-2)}{M_- N_+} \right) \nonumber \\
&\rightarrow& \left(\frac{\sqrt{-1}(L-1)}{2K}, \frac{-(L+1)}{2K} \right) \nonumber \\
&\rightarrow& \left(\frac{3x+1+\sqrt{-1}u}{M_+ N_-}, \frac{-u+\sqrt{-1}(L+x+2)}{M_+ N_-} \right) \nonumber \\
&\rightarrow& \left(\frac{\sqrt{-1}(L+2x-3)}{2N_+}, \frac{L+3+2\sqrt{-1}u}{N_+} \right) \nonumber \\
&\rightarrow& \left(\frac{\sqrt{-1}(L-2x-1)}{2M_-}, \frac{-L-1-2\sqrt{-1}u}{2M_-} \right) \nonumber \\
&\rightarrow& \left(\frac{L+4x-1+2\sqrt{-1}u}{2M_+K}, \frac{-4u+\sqrt{-1}(L+2x+1)}{2M_+K} \right) \nonumber \\
&\rightarrow& \left(\frac{-L+x+2+\sqrt{-1}u}{N_+ N_-}, \frac{-3u+\sqrt{-1}(x+1)}{N_+ N_-} \right) \nonumber \\
&\rightarrow& \left(x, u \right) \nonumber
\end{eqnarray}
where $M_{\pm}=-\sqrt{-1}(1-x)\pm u$, $N_{\pm}=\sqrt{-1}(1+x)\pm u$, $K=-u+\sqrt{-1}x$, $L=x^2+u^2$.
\end{document}